\documentclass{article}
\usepackage{spconf}
\usepackage{amsmath,graphicx}
\usepackage{amssymb}
\usepackage{amsmath}
\usepackage{algorithm}
\usepackage{algpseudocode}
\usepackage{pgfplots, pgfplotstable}
\usepackage[utf8x]{inputenc}
\usepackage{booktabs}
\usepackage{microtype}
\usepackage{flushend}
\graphicspath{{figs/}}

\title{Hardware Implementation of Iterative Projection-Aggregation Decoding of Reed-Muller Codes}
\name{Marzieh Hashemipour-Nazari, Kees Goossens, Alexios Balatsoukas-Stimming}
\address{}
\begin{document}
%
\maketitle
\begin{abstract}
In this work, we present a simplification and a corresponding hardware architecture for hard-decision recursive projection-aggregation (RPA) decoding of Reed-Muller (RM) codes. In particular, we transform the recursive structure of RPA decoding into a simpler and iterative structure with minimal error-correction degradation.
Our simulation results for RM$(7,3)$ show that the proposed simplification has a small error-correcting performance degradation ($0.005$ in terms of channel crossover probability)  while reducing the average number of computations by up to 40\%.
In addition, we describe the first fully parallel hardware architecture for simplified RPA decoding. We present  FPGA implementation results for an RM$(6,3)$ code on a Xilinx  Virtex-7 FPGA showing that our proposed architecture achieves a throughput of 171 Mbps at a frequency of 80 MHz.   
\end{abstract}
%
%
\section{Introduction}
\label{sec:intro}
Reed-Muller (RM) codes were first proposed in 1954 \cite{Reed1954}. Recently, there has been a renewed interest in RM codes because, in some cases, they were shown to be capable of achieving the Shannon capacity of the binary erasure channel (BEC) \cite{Kudekar2017} and the binary symmetric channel (BSC) \cite{Sberlo2020}.
The oldest decoding algorithm for RM codes is based on majority voting \cite{Reed1954}, and it guarantees correction of the error patterns with a weight less than half of the minimum distance. A wide variety of algorithms has been proposed afterward to improve decoding capacity. For example, the Sidel’nikov-Pershakov algorithm \cite{Sidel1992} corrects most of the corrupted codewords with a number of errors less than $(1-\varepsilon)n/2$, where $n$ indicates blocklength of the RM codes and $\varepsilon\geq n^{{-}1/3}$. Some hardware architectures are also available for the aforementioned methods. A parallel decoding architecture for majority-logic decoding algorithms was provided in \cite{Bertram2013}, and a low-area decoder for the Reed decoding method was introduced in \cite{Hiller2015}. 

Successive-cancellation (SC) decoding \cite{Dumer2004} and SC list (SCL) decoding \cite{Dumer2006}, make use of the decomposable structure of RM codes to provide recursive decoding methods with reasonable complexity.
The work of \cite{Hashemi2018} improved the performance of SC and SCL decoding methods by exploring several carefully selected permutations of the factor graph of RM codes.
The work of \cite{Santi2018} exploits the symmetric structure of RM codes and applies an iterative decoding method to provide near maximum likelihood (ML) performance.
Other works focused on special cases of RM codes. For example, \cite{Sakkour2005} is a modified version of the Sidel’nikov-Pershakov algorithm that improves error-correcting performance for second-order RM codes. Moreover, the work of \cite{Thangaraj2020} provided a new ML decoder with a lower complexity for RM codes of order $m-3$, where $m=\log_2 n$. 

The main drawback of the aforementioned algorithms is that they have poor error-correcting performance for short blocklength RM codes. For this reason, the authors of~\cite{Ye2020} proposed a new algorithm called recursive projection-aggregation (RPA) decoding that improves the error-correcting performance of RM codes in the regimes of interest of ultra-reliable low-latency communications (URLLC) and of the Internet of Things (IoT), i.e., low rate and short blocklength RM codes. 
The RPA algorithm is highly parallelizable. However, it has a high complexity and its recursive structure is not particularly amenable to hardware implementations.

\emph{Contributions:} In this paper, we present a simplified version of the RPA algorithm to make a trade-off between the error-correcting performance and computations. 
We simplify the RPA algorithm by carefully removing computations in the recursion levels to make the structure suitable for hardware implementations.
Moreover, we propose the first fully parallel hardware architecture for RPA decoding. 
%
%

\section{Reed-Muller Codes}
\label{sec:RMcode}

The focus of this paper is on the BSC, so all operations and vectors are in $\mathbb{F}_{2}$.
RM codes are denoted by $\mathrm{RM}(m,r)$, where $m$ indicates the code length $n=2^m$ and $r$ is the order.
RM codes are linear block codes with rate ${R=\frac{k}{n}}$, ${k=\sum _{i=0}^r \binom{n}{i}}$, and with the following recursively defined generator matrix: 
\begin{equation}\label{eq:GMtx}
\mathbf{G}_{(m, r)}= \begin{bmatrix}
  \mathbf{G}_{(m-1, r)} & \mathbf{G}_{(m-1, r)}\\ 
  \mathbf{0} & \mathbf{G}_{(m-1, r-1)}
\end{bmatrix},
\quad
\mathbf{G}_{(1, 1)}{=} \begin{bmatrix}
  1 & 1\\ 
  0 & 1
\end{bmatrix}.
\end{equation}

\subsection{Recursive Aggregation Projection Decoding}
\label{Sec:RPA}
\begin{algorithm}[t]
\footnotesize
\caption{The {\ttfamily{RPA}} decoding of RM codes RM$(m,r)$}
\label{alg:RPA}
\textbf{Input}: The noisy codeword $\mathbf{y}$, $m$, $r$, $N_{\max}$\\
\textbf{Output}:  The decoded codeword $\mathbf{c}$ \\
1: \textbf{if} {$r=1$} \textbf{do}\\
2: \mbox{\hspace{2mm} $c \leftarrow${\ttfamily{order-1-decoding}} $(\mathbf{y},m)$ }\\
3: \textbf{else}\\
4: \mbox{\hspace{2mm} \textbf{for} $j=1:N_{\max}$ \textbf{do}}\\
5: \mbox{\hspace{10mm} \textbf{for} $i=1:2^m-1$ \textbf{do}}\\
6: \mbox{\hspace{15mm} $\mathbf{y}_{i} \leftarrow$ {\ttfamily{Proj}}$(\mathbf{y},i,m)$}\\
7: \mbox{\hspace{15mm} $\hat{\mathbf{y}}_{i} \leftarrow$ {\ttfamily{RPA}}$(\mathbf{y}_{i},m-1,r-1,N_{\max})$}\\
8: \mbox{\hspace{10mm} \textbf{end for}}\\
9:\mbox{\hspace{10mm} $\hat{\mathbf{y}} \leftarrow$ {\ttfamily{Agg}}$(\mathbf{y},\mathbf{y}_1,\mathbf{y}_2,\ldots,\mathbf{y}_{n-1},\hat{\mathbf{y}}_2,\hat{\mathbf{y}}_2,\ldots,\hat{\mathbf{y}}_{n-1})$}\\
10:\mbox{\hspace{10mm}\textbf{if} $\mathbf{y} = \hat{\mathbf{y}}$ \textbf{do} }\\
11:\mbox{\hspace{15mm}\textbf{break} \hspace{8mm}  \texttt{-{}-}\textit{RPA converges to a fixed point }}\\
12:\mbox{\hspace{10mm}\textbf{end if}}\\
13:\mbox{\hspace{10mm}$\mathbf{y} \leftarrow \hat{\mathbf{y}}$}\\
14: \mbox{\hspace{2mm}\textbf{end for}}\\
15: \mbox{\textbf{end if}}
\end{algorithm}
\begin{algorithm}[t]
\footnotesize
\caption{The projection function {\ttfamily{Proj}}}
\label{alg:Proj}
\textbf{Input}: $\mathbf{y}_\text{in}(0 \; \text{to} \;n{-}1), i, m $ \\
\textbf{Output}: $\mathbf{y}_\text{out}(0 \; \text{to} \;n/2-1)$ \\
1: $n \leftarrow 2^m$\\
2: \textbf{if} {$i < n/2$} \textbf{do}\\
3: \mbox{\hspace{2mm}$\mathbf{y}_\text{out}(0 \; \text{to} \;n/4{-}1)\leftarrow ${\ttfamily{Proj}}$( \mathbf{y}_\text{in}(0 \; \text{to} \;n/2{-}1),m{-}1,i)$}\\
4: \mbox{\hspace{2mm}$\mathbf{y}_\text{out}(n/4 \; \text{to} \;n/2{-}1) \leftarrow ${\ttfamily{Proj}}$(\mathbf{y}_\text{in}(n/2 \; \text{to} \;n{-}1),m{-}1,i)$}\\
5: \textbf{else}\\
6: \mbox{\hspace{5mm} \textbf{for} $j{=}1$ : $n/2{-}1$ \textbf{do}}\\
7: \mbox{\hspace{10mm}$\mathbf{y}_\text{tmp}(2j) \leftarrow \mathbf{y}_\text{in}(j)$}\\
8: \mbox{\hspace{10mm}$\mathbf{y}_\text{tmp}(2j{+}1) \leftarrow \mathbf{y}_\text{in}($bi2de(de2bi$(j) \oplus $de2bi$(i))$}\\
9: \mbox{\hspace{5mm} \textbf{end for}}\\
10:  \mbox{\hspace{5mm}$\mathbf{y}_\text{tmp}(0) \leftarrow \mathbf{y}_\text{in}(0)$}\\
11: \mbox{\hspace{5mm}$\mathbf{y}_\text{tmp}(1) \leftarrow \mathbf{y}_\text{in}(i)$}\\
12: \mbox{\hspace{5mm}\textbf{for} $t{=}0$ : $n/2{-}1$ \textbf{do}}\\
13: \mbox{\hspace{10mm}$\mathbf{y}_\text{out}(t) \leftarrow \mathbf{y}_\text{tmp}(2t) \oplus \mathbf{y}_\text{tmp}(2t+1)$}\\
14: \mbox{\hspace{5mm}\textbf{end for}}\\
15: \textbf{end if}
\end{algorithm} 
\begin{algorithm}[t]
\footnotesize
\caption{The aggregation function {\ttfamily{Agg}}}
\label{alg:Agg}
\textbf{Input}: $m, \mathbf{y}_\text{in},\mathbf{y}_{1},\mathbf{y}_{2} {\ldots},\mathbf{y}_{n-1}, \hat{\mathbf{y}}_{1}, \hat{\mathbf{y}}_{2},{\ldots},\hat{\mathbf{y}}_{n-1} $ \\
\textbf{Output}: $\mathbf{\mathbf{y}}_\text{out}$ \\
1:\hspace{1mm}\textbf{for} $z=0:2^{m}-1$ \textbf{do}\\
2:\mbox{\hspace{5mm} $vote(z) \leftarrow 0$}\\
3:\mbox{\hspace{5mm} \textbf{for} $i=1:2^{m}-1$ \textbf{do}}\\
4:\mbox{\hspace{10mm} $Ind \leftarrow $ {\ttfamily{FindIndex}}$(z,i,m)$ }
5:\mbox{\hspace{10mm} $vote(z) \leftarrow \mathbf{y}_{i}(Ind) \oplus \hat{\mathbf{y}}_{i}(Ind) + vote(z)$}\\
6:\mbox{\hspace{5mm} \textbf{end for}}\\
7:\mbox{\hspace{5mm} $\mathbf{y}_\text{out}(z) \leftarrow \mathbf{y}_\text{in}(z) \oplus \mathbf{1} \left[vote(z)> \frac{2^m-1}{2}\right] $}\\
8:\hspace{1mm}\textbf{end for}
\end{algorithm}

\begin{algorithm}[t]
\footnotesize
\caption{The function {\ttfamily{FindIndex}}}
\label{alg:FinInd}
\textbf{Input}: Index $z$, branch number $i$, $m$ \\
\textbf{Output}: $Ind$ \\
1:\hspace{1.3mm}\textbf{if} $i\geq 2^{m-1}$ \textbf{do}\\
2:\mbox{\hspace{5mm}\textbf{if} $z<2^{m-1}$ \textbf{do}}\\
3:\mbox{\hspace{10mm}$Ind \leftarrow z$}\\
4:\mbox{\hspace{5mm}\textbf{else}}\\
5:\mbox{\hspace{10mm}$Ind \leftarrow$ bi2de(de2bi $(z) \oplus$ de2bi$(i))$}\\
6:\mbox{\hspace{5mm}\textbf{end if}}\\
7:\hspace{1.5mm}\textbf{else} \\
8:\mbox{\hspace{6mm}\textbf{if} $z<2^{m-1}$ \textbf{do}}\\
9:\mbox{\hspace{10mm}$Ind \leftarrow$ {\ttfamily{FindIndex}}$(z,i,m-1)$}\\
10:\mbox{\hspace{5mm}\textbf{else}}\\
11:\mbox{\hspace{8.7mm}$Ind \leftarrow$ {\ttfamily{FindIndex}}$(z-(2^{m-1}),i,m-1) + 2^{m-2}$}\\
12:\mbox{\hspace{5mm}\textbf{end if}}\\
13:\hspace{1mm}\textbf{end if}
\end{algorithm}

As Algorithm~\ref{alg:RPA} shows, the RPA algorithm has three main steps:  projection (line 6), recursive decoding (line 7), and aggregation (line 9). Let us consider a noisy received vector  $\mathbf{y}$ of the transmitted codeword  $\mathbf{c}$ of length $n$. 

In the \textit{projection} step, $\mathbf{y}$ is transformed into $n-1$ distinct vectors of length $n/2$. For hard-decision decoding, each transformed vector $\mathbf{y}_i, i\in\{1,2,{\ldots},n{-}1\}$, is obtained by taking the modulo-2 sum over specific coordinates of the input vector corresponding to the $i$-th projection, as shown on lines 2-11 of Algorithm~\ref{alg:Proj}. Next, on line 13, a binary \emph{XOR} operation sums every two adjacent bits to convert each $n$-bit input vector $\mathbf{y}$ to an $n/2$-bit vector $\mathbf{y}_{i}$. 

In the \textit{recursive decoding} step, each vector $\mathbf{y}_i$, produced in the projection step, is recursively decoded by RPA for RM$(m{-}1,r{-}1)$ until first-order RM codes are reached, which can be decoded efficiently using the fast Hadamard transform (FHT) \cite{Beext86}. Each $\hat{\mathbf{y}}_i$ is a decoded vector of $\mathbf{y}_i$.

In the \textit{aggregation} step, for each coordinate $\hat{\mathbf{y}}(z)$, Algorithm~\ref{alg:FinInd} finds the corresponding coordinates in $\mathbf{y}_i$ that were originally created with $\mathbf{y}(z)$. These coordinates together with their decoded value in $\hat{\mathbf{y}}_i$ represent $n-1$ estimations for each coordinate of vector $\hat{\mathbf{y}}$ (see line 3-6 in Algorithm~\ref{alg:Agg}).
Next, in line 7 of Algorithm~\ref{alg:Agg}, per-coordinate majority voting is performed to produce an estimate $\hat{\mathbf{y}}$ of the transmitted codeword.

This procedure is repeated for multiple iterations. In \cite{Ye2020}, the maximum number of iterations is set to $N_{\max}=\lceil m/2 \rceil$.

\section{Iterative Projection-Aggregation Decoding}
\begin{algorithm}[t]
\footnotesize
\caption{ The {\ttfamily{IPA}} decoding of RM$(m,r)$ codes }
\label{alg:SRPA}
\textbf{Input}: The noisy codeword $\mathbf{y}$, $m$, $r$, $N_{\max}$\\
\textbf{Output}:  The decoded codeword $\mathbf{c}$ \\
1:\mbox{\hspace{2mm}$\mathbf{y}_{(1,0)} \leftarrow \mathbf{y}; \hspace{5mm} n_{tmp}\leftarrow 1$}\\
2:\hspace{2mm}\textbf{for} $j=1:N_{\max}$ \textbf{do}\\
3:\mbox{\hspace{6mm}\textbf{for} $l=1:r-1$ \textbf{do}  \hspace{15mm} \texttt{-{}-}\textit{Projection loop}}\\
4:\mbox{\hspace{11mm} $n_{tmp} \leftarrow n_{tmp}\times (2^{m}-1)$}\\
5:\mbox{\hspace{11mm}\textbf{for} $i=1:n_{tmp}$ \textbf{do}}\\
6:\mbox{\hspace{15mm} $\mathbf{y}_{tmp} \leftarrow \mathbf{y}_{(\lceil{i/(2^m-1)\rceil},l-1)}$}\\
7:\mbox{\hspace{15.5mm}$\mathbf{y}_{(i,l)} \leftarrow ${\ttfamily{Proj}}$(\mathbf{y}_{tmp},mod(i,2^m-1),m)$}\\
8:\mbox{\hspace{11mm}\textbf{end for}}\\
9:\mbox{\hspace{11mm}$m \leftarrow m-1$}\\
10:\mbox{\hspace{6mm}\textbf{end for}}\\
11:\mbox{\hspace{6mm}\textbf{for} $t:1:n_{tmp}$ \textbf{do} \hspace{15mm} \texttt{-{}-}\textit{First-order decoding}}\\
12:\mbox{\hspace{10.5mm} $\hat{\mathbf{y}}_{(t,r-1)} \leftarrow$ {\ttfamily{order-1-decoding}} $(\mathbf{y}_{(t,r-1)},m)$}\\
13:\mbox{\hspace{6mm}\textbf{end for}}\\
14:\mbox{\hspace{6mm}\textbf{for} $l=r-2:0$  \textbf{do} \hspace{15mm} \texttt{-{}-}\textit{Aggregation loop}}\\
15:\mbox{\hspace{11mm} $m \leftarrow m+1;  \hspace{5mm} t \leftarrow 2^m-1; \hspace{5mm} n_{tmp} \leftarrow n_{tmp}/t$  }\\
16:\mbox{\hspace{11mm}\textbf{for} $i=1:n_{tmp}$ \textbf{do}}\\
17:\mbox{\hspace{15mm} $d	\leftarrow (i-1)\times t $}\\
18:\mbox{\hspace{15.5mm}$\mathbf{\hat{y}}_{(i,l)} \leftarrow ${\ttfamily{Agg}}$(\mathbf{y}_{(i,l)}, \mathbf{y}_{(d+1,l+1)},{\ldots},\mathbf{y}_{(d+t,l+1)}, $}\\
\mbox{\hspace{37mm}$\mathbf{\hat{y}}_{(d+1,l+1)},{\ldots},\mathbf{\hat{y}}_{(d+t,l+1)})$}\\
19:\mbox{\hspace{11mm}\textbf{end for}}\\
20:\mbox{\hspace{6mm}\textbf{end for}} \\
21:\mbox{\hspace{6mm}\textbf{if} $\mathbf{y}_{(1,0)} = \hat{\mathbf{y}}_{(1,0)}$ \textbf{do} }\\
22:\mbox{\hspace{11mm}\textbf{break} \hspace{8mm}  \texttt{-{}-}\textit{{\ttfamily{IPA}} converges to a fixed point } }\\
23:\mbox{\hspace{6mm}\textbf{end if}}\\
24:\mbox{\hspace{5.5mm}$\mathbf{y}_{(1,0)} \leftarrow \hat{\mathbf{y}}_{(1,0)}$}\\
25:\hspace{2mm}\textbf{end for}\\
26:\mbox{\hspace{2mm}$\mathbf{c} \leftarrow \hat{\mathbf{y}}_{(0,1)}$}

\end{algorithm}

As can be seen on line 4 of Algorithm~\ref{alg:RPA}, RPA decoding performs multiple iterations at each level of the recursion. 
After each aggregation, if $\hat{\mathbf{y}} \neq \mathbf{y}$, $\mathbf{y}$ will be updated by $\hat{\mathbf{y}}$, and the whole procedure from projection to aggregation will iterate again.
Unfortunately, having iterations on each recursion level makes the RPA structure complicated, especially for hardware implementations,  as it requires very complicated control circuitry and memory structures. 

In the case of a noisy received vector $\mathbf{y}$ with only one error, for all projected vectors at each recursion level, it can be verified from Algorithm~\ref{alg:Proj} that there exists exactly one error for every level of the recursion and for all projections, which is corrected in level $r=1$ because FHT decoding guarantees the correction of one error. 
However, the condition for skipping the remaining iterations is not satisfied (see line 10 in Algorithm~\ref{alg:RPA}), and as a result, RPA runs another iteration at this level. 
This additional iteration is unnecessary because it performs projection, first-order decoding, and aggregation on the already corrected codewords. 
More generally, and motivated by the above example, if the iteration loops for a recursion level run more than once but stop before reaching $N_{\max}$, the last iteration always runs only to check the stop condition. We call these iterations \emph{ineffective}.

\begin{figure*}[t]
  \centering
  \centerline{\includegraphics[width=0.85\textwidth]{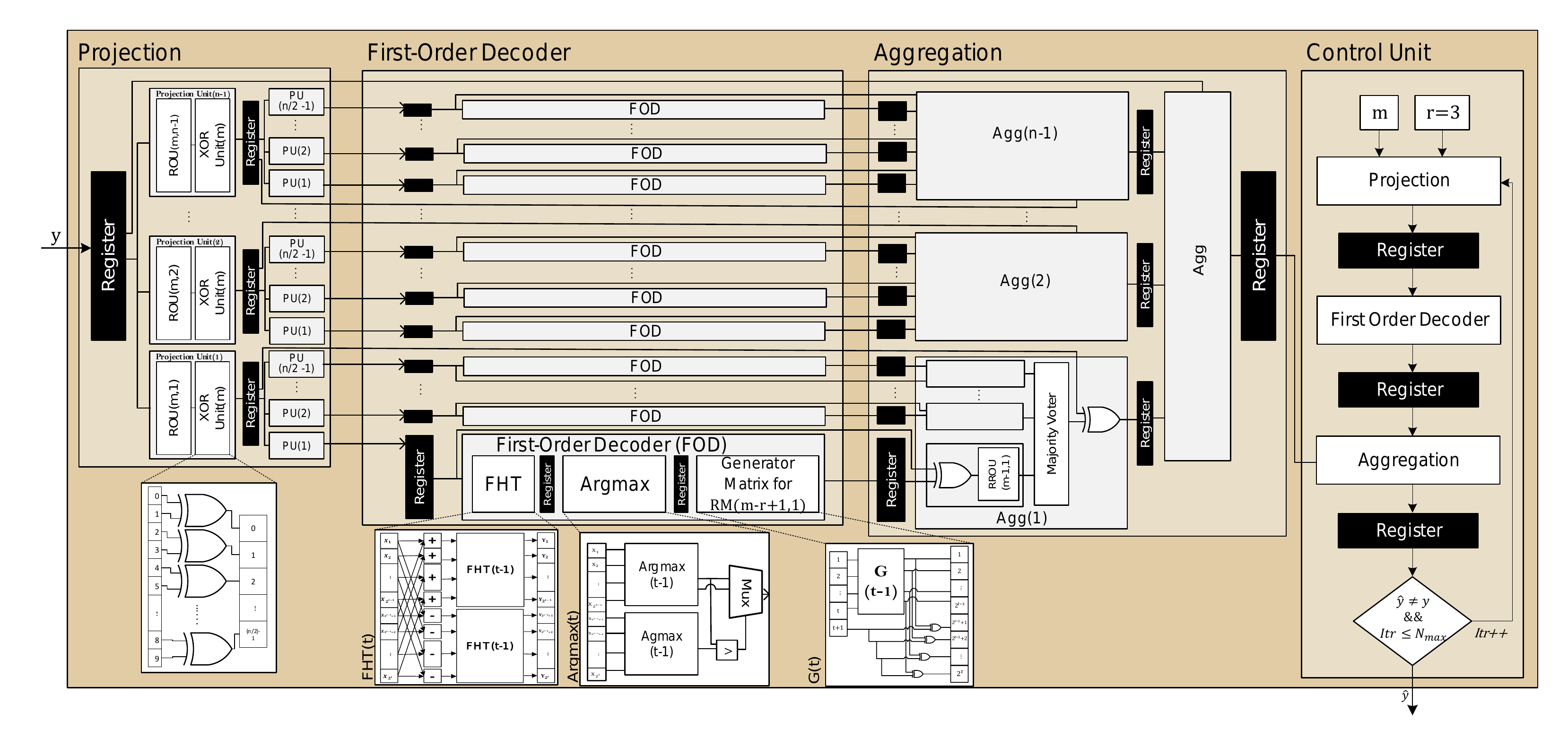}}
  \vspace{-0.6cm}
  \caption{An example of the proposed hardware architecture for IPA decoding for RM$(m,3)$ codes.}
  \label{fig:Arch}
  \vspace{-0.1cm}
\end{figure*} 

Based on our simulations of various RM codes, at low channel crossover probabilities, more than $50\%$ of internal iterations are ineffective. Motivated by this observation, we present a simplification of RPA by removing iterations on the internal levels of the RPA recursion. Effectively, our proposed iterative projection-aggregation (IPA) algorithm sets $N_{\max}=1$ for all recursive decoding steps of the RPA algorithm except for the first one. Moreover, in Algorithm~\ref{alg:SRPA}, we show an iterative structure that is more convenient for a hardware implementation. As we show in Section~\ref{sec:result}, this reduces complexity and hardware implementation significantly, with only a small penalty in the error-correcting performance.

It can be shown that the complexity of RPA decoding with internal iterations is $O(n^r (\log_2 n)^{r+1})$. For the IPA algorithm, the complexity is $O(n^r(\log_2 n)^2)$ as we remove the internal iterations. We also show in the Section~\ref{sec:result} that the overall calls to the first-order decoder, which is a more practical complexity measure, are decreased significantly.
 

\section{Proposed Hardware Architecture}
\label{sec:HW}

Our proposed fully parallel IPA architecture, which is shown in Fig.~\ref{fig:Arch}, consists of three main components and a control unit. The first component is the \emph{projection}, including $r-1$ levels of the projection for RM$(m, r)$ codes (line 3 of Algorithm~\ref{alg:SRPA}). The second component, which we call the \emph{first-order decoder}, has parallel decoders for all RM$(m-r+1,1)$ codes generated in the innermost level of the RPA (line 11 of Algorithm~\ref{alg:SRPA}). The third component is the \emph{aggregation} unit performing $r-1$ levels of aggregation (line 14 of Algorithm~\ref{alg:SRPA}).
 
The \emph{projection} component performs {$r-1$} levels of projection, as described in Section~\ref{Sec:RPA}. Each projection level has parallel projection units, each consisting of a \emph{re-ordering} unit (\emph{ROU}) and an \emph{XOR} unit. The re-ordering unit \emph{ROU(m,i)} finds the coordinates for $i$-th projection of the input vector $\mathbf{y}$ with length of $2^m$  based on lines 2-11 of Algorithm~\ref{alg:Proj}. Then, an \emph{XOR} unit is assigned to each projection branch for performing the sum operations as described in lines 12-13 of Algorithm~\ref{alg:Proj}.

The \emph{first-order decoder} component provides first-order decoders (FODs) for all RM$(m{-}r{+}1,1)$ codes,  obtained in the innermost level of projection, in parallel. Each \emph{FOD} was designed based on the decoding method proposed in \cite{Beext86}, and consists of three sub-units: \emph{FHT}, \emph{Argmax}, and \emph{Generator matrix}.  
The first unit gives the vector $\mathbf{l}$, which is the result of the FHT on a binary input vector $\mathbf{y}$: 
\begin{equation} \label{eq:fht}
\mathbf{l}{=}(1-2\mathbf{y})\mathbf{H}_{2^m},
\end{equation}
where the Hadamard matrix $\mathbf{H}_{2^m}$ is 
\begin{equation} \label{eq:fhtmatrix}
\mathbf{H}_{2^m} = \begin{bmatrix}
  \mathbf{H}_{2^{m-1}} & \mathbf{H}_{2^{m-1}}\\ 
  \mathbf{H}_{2^{m-1}} & -\mathbf{H}_{2^{m-1}} 
\end{bmatrix} \text{ and }
\: \: \mathbf{H}_2 = \begin{bmatrix}
1 & 1\\
1 & -1
\end{bmatrix}.
\end{equation}
The architecture of \emph{FHT} unit is derived from \cite{Agrawal2015}. 
The \emph{Argmax} unit finds the index $z$ of the maximum value of $\mathbf{l}$. The output of the \emph{FOD} unit is:
\begin{align}
	\hat{\mathbf{y}} & = \hat{\mathbf{x}}\mathbf{G}_{(m, 1)},
\end{align}
where $\hat{\mathbf{x}}=\begin{bmatrix} \frac{1-\text{sign}(\mathbf{l}(z))}{2} & \mathbf{z}_{\text{bin}} \end{bmatrix}$ with $\mathbf{z}_{\text{bin}}$ being the binary representation of $z$, and where $\mathbf{G}_{(m, 1)}$ is the generator matrix of RM$(m,1)$.

The \emph{aggregation} component provides $r{-}1$ levels of aggregation, each of which has
$\frac{\prod_{i=0}^{r{-}2} (2^{m{-}i}-1)}{\prod_{i=1}^{j}(2^{(m{-}r{+}1){+}i}-1)}$ \emph{AGG} units in parallel (line 16 of Algorithm~\ref{alg:SRPA}), where $j$ denotes the current level of aggregation. As Fig. \ref{fig:Arch} shows, each \emph{AGG} unit consists of  $2^m{-}1$ \emph{RRUM}s (the hardware implementation of Algorithm~\ref{alg:FinInd}) and one majority voter to aggregate into a $n$-bit codeword $\hat{\mathbf{y}}$ calculated in line 7 of Algorithm~\ref{alg:Agg}. 
Finally, XOR gates are used to flip the desired bits of input vector $\mathbf{y}$ in $\hat{\mathbf{y}}$ as described in line 7 of Algorithm~\ref{alg:Agg}.

\begin{figure}[t]
\centering
\begin{tikzpicture}
\begin{semilogyaxis}[
    height=0.65 \columnwidth,
    width= 0.9\columnwidth,
    yminorgrids = true,
    xminorgrids = true,
	xlabel=Channel crossover probability,
	ylabel=Frame error rate,
	legend pos=south east,
    ymajorgrids=true,
    xmajorgrids=true,
    xticklabel style={/pgf/number format/fixed},
    grid style=dashed,
    xmin = 0,
	xmax = 0.1,
	ymax = 1,
	ymin = 1e-5,
	legend style={nodes={scale=0.625, transform shape}}]

\addplot[ color=red,mark=square*] coordinates {
    (0.00425,0.00018)
    (0.0055,0.00036)
    (0.00675,0.00120)
    (0.008,0.0023)
    (0.00925,0.0038)
    (0.0105,0.0058)
    (0.01175,0.0076)
    (0.013,0.0103)
    (0.01425,0.0127)
    (0.0155,0.0167)
    (0.018,0.0252)
    (0.023,0.0522)
    (0.028,0.0925)
    (0.033,0.1490)
    (0.043,0.2940)
    (0.06,0.546)
};
\addplot[ color=blue,mark=*] coordinates {
    (0.00425,1.8000e-04)
    (0.0055,0.00036)
    (0.00675,0.0012)
    (0.008,0.0023)
    (0.00925,0.0038)
    (0.0105,0.0058)
    (0.01175,0.0076)
    (0.013,0.0103)
    (0.01425,0.0127)
    (0.0155,0.0167)
    (0.018,0.0252)
    (0.023,0.0522)
    (0.028,0.0925)
    (0.033,0.1490)
    (0.043,0.2940)
    (0.06,0.546)
};

\addplot[dashed,color=red,mark=square*,mark options={solid}]
 coordinates {
    (0.025,0.00004)
    (0.03,0.00033)
    (0.035,0.0013)
    (0.04,0.0036)
    (0.045,0.0076)
    (0.05,0.0134)
    (0.055,0.0250)
    (0.06,0.0410)
    (0.065,0.068)
    (0.07,0.106)
};
\addplot[dashed, color=blue,mark=*,mark options={solid}]
 coordinates {
    (0.025,7.0e-05)
    (0.03,0.00044)
    (0.035,0.0021)
    (0.04,0.0047)
    (0.045,0.010)
    (0.05,0.0185)
    (0.055,0.0360)
    (0.06,0.070)
    (0.065,0.102)
    (0.07,0.148)
};
\addplot[dashed, color=black,mark=x,mark options={solid, scale=1.5}]
 coordinates {
    (0.025,7.0e-05)
    (0.03,0.00044)
    (0.035,0.0021)
    (0.04,0.0047)
    (0.045,0.010)
    (0.05,0.0185)
    (0.055,0.0360)
    (0.06,0.070)
    (0.065,0.102)
    (0.07,0.148)
};

\legend{RPA RM{$(6,3)$},IPA RM{$(6,3)$}, RPA RM{$(7,3)$}, IPA RM{$(7,3)$},IPA RM{$(7,3)$}\, $N_{\max}{=}20$}
\end{semilogyaxis}
\end{tikzpicture}
\caption{Frame error rate Comparison between RPA and IPA for RM$(6,3)$ and RM$(7,3)$ codes over BSC channels.}
\label{fig:FER}
\vspace{-0.1cm}
\end{figure}
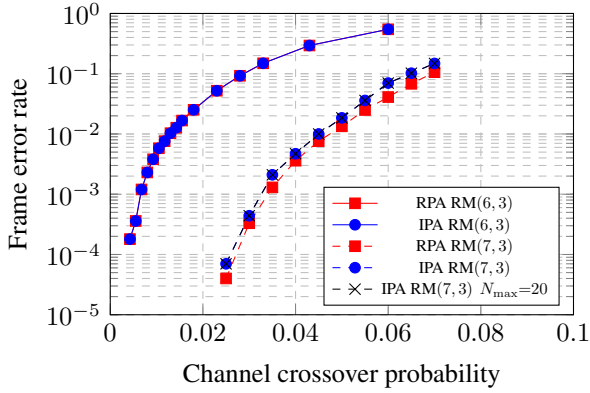

The throughput of the decoder is calculated by:
\begin{equation}
\text{Throughput} = \frac{\text{Frequency}}{N_{\text{iter}}N_{\text{cycles/iter}}} \times n,
\end{equation}
where $N_{\text{iter}} = N_{\max}$ for the minimum throughput and $N_{\text{iter}} = N_{\text{avg}}$ is the average number of iterations for the average throughput. Registers are inserted to the architecture to reduce the critical path through the proposed decoder. In particular, $r-1$ and $2(r-1)$ registers are inserted between the projection and aggregation levels, respectively. Additionally, three registers are inserted between the components of the FODs, and one register is used to check the termination condition. As such, we have $N_{\text{cycles/iter}} = 3(r-1)+4$.

\section{Results}
\label{sec:result}

Simulation results for IPA decoding and RPA decoding for the RM$(6,3)$ and RM$(7,3)$ codes over the BSC channel are shown in Fig.~\ref{fig:FER}. We observe that IPA decoding has exactly the same frame error rate (FER) as RPA decoding for RM$(6,3)$, while there is a minimal error-correcting performance degradation up to $0.005$ in terms of channel cross-over probability for RM$(7,3)$. We also increased $N_{\max}$ to see if this compensates the performance degradation of IPA, but we observed that it unfortunately does not help.

\begin{figure}[t]
\centering
\begin{tikzpicture}
\begin{axis}[
    height=0.7 \columnwidth,
    width= 0.9\columnwidth,
    yminorgrids = true,
    xminorgrids = true,
	xlabel=Channel crossover probability,
	ylabel=Avg. number of order-1 decodings ,
	legend pos=north west,
    ymajorgrids=true,
    xmajorgrids=true,
    xmin = 0,
    xmax = 0.07,
    xticklabel style={/pgf/number format/fixed},
    grid style=dashed,
	legend style={nodes={scale=0.625, transform shape}}]
\addplot[color=red,mark=square*] coordinates {
    (0.00425,2868.38)
    (0.0055,3097.89)
    (0.00675,3299.46)
    (0.008,3491.502)
    (0.00925,3661.217)
    (0.0105,3822.93)
    (0.01175,3993.894)
    (0.013,4122.9)
    (0.01425,4248.67)
    (0.0155,4359.25)
    (0.018,4558.9)
    (0.023,4837.7)
    (0.028, 4984.4)
    (0.033,5036.3)
    (0.043,5137.3)
    (0.06,5310)
};
\addplot[color=blue,mark=*,mark options={solid}] coordinates {
    (0.00425,2411)
    (0.0055,2525.9)
    (0.00675,2626.4)
    (0.008,2721.9)
    (0.00925,2806.1)
    (0.0105,2886.1)
    (0.01175,2970.7)
    (0.013,3034)
    (0.01425,3096.1)
    (0.0155,3149.2)
    (0.018,3244.5)
    (0.023,3368.1)
    (0.028, 3418.3)
    (0.033,3428.5)
    (0.043,3465.4)
    (0.06,3470)
};
\addplot[dashed, color=red,mark=square*,mark options={solid}] coordinates {
    (0.025,23364)
    (0.03,23689)
    (0.035,23921)
    (0.04,24105)
    (0.045,24296)
    (0.05,24622)
    (0.055,25226)
    (0.06,25820)
    (0.065,26518)
    (0.07,27320)
};
\addplot[dashed, color=blue,mark=*,mark options={solid}] coordinates {
    (0.025,15700)
    (0.03,15800)
    (0.035,15970)
    (0.04,16056)
    (0.045,16175)
    (0.05,16334)
    (0.055,16546)
    (0.06, 16754)
    (0.065,16890)
    (0.07,16900)
};
\legend{RPA RM{$(6,3)$},IPA RM{$(6,3)$}, RPA RM{$(7,3)$},IPA RM{$(7,3)$}}
\end{axis}
                                                             
\end{tikzpicture}
\caption{Comparison of the number of the first-order decoding between RPA and IPA for RM$(6,3)$ and RM$(7,3)$ codes.}
\label{fig:FHT}
\vspace{-0.1cm}
\end{figure}
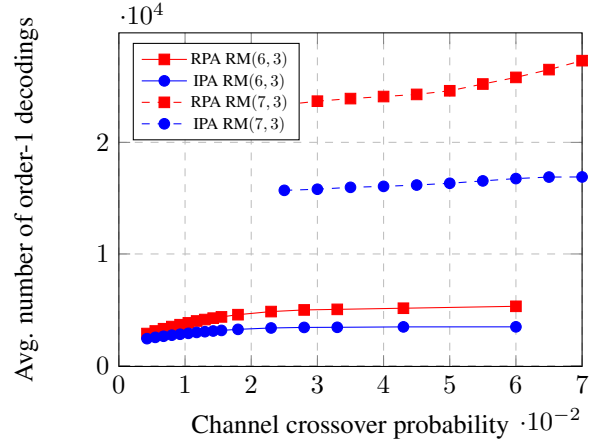

\begin{table}[t]
\centering
\caption{Post-PAR results for an RM$(6,3)$ code on a Xilinx Virtex-7 FPGA (xc7vx1140T).}
\small
\begin{tabular}{ll} 
 \toprule
 LUTs    			& $602,111/712,000 (84.57\%)$   \\
 Flip-flops  		& $65,699/1,424,000 (4.6\%)$  \\
 Clock frequency 	& $80$ MHz \\
 Min. throughput    & $171$ Mbps \\
 Avg. throughput @ FER=$10^{-3}$    &$284$ Mbps \\
 \bottomrule
\end{tabular}
\label{Table:synth}
\end{table}

Fig.~\ref{fig:FHT} shows the average number of first-order decodings for IPA and RPA decoding for RM$(6,3)$ and RM$(7,3)$ codes over the BSC channel. The number of the first-order decodings is decreased by up to $40\%$ for RM$(6,3)$ and up to $50\%$ for RM$(7,3)$.
Therefore, our proposed IPA decoding provides a good trade-off between FER and computations.

We provide post-PAR results of our IPA decoder architecture for RM$(6,3)$ on a Xilinx Virtex-7 FPGA with a frequency of $80$~MHz in Table~\ref{Table:synth}. The resource utilization is high due to the fully parallel nature of the decoder, but the achieved decoding throughput is also relatively high. There are no other implementations of RPA in the literature, we cannot perform a direct comparison.



%


%
%
\bibliographystyle{IEEEbib}
\bibliography{refs}

\begin{thebibliography}{10}

\bibitem{Reed1954}
I.~Reed,
\newblock ``A class of multiple-error-correcting codes and the decoding
  scheme,''
\newblock {\em Transactions of the {IRE} Professional Group on Information
  Theory}, vol. 4, no. 4, pp. 38--49, sep 1954.

\bibitem{Kudekar2017}
S.~Kudekar, S.~Kumar, M.~Mondelli, H.~D. Pfister, E.~\c{S}a\c{s}o\u{g}lu, and
  R.~Urbanke,
\newblock ``{Reed{\textendash}Muller} codes achieve capacity on erasure
  channels,''
\newblock {\em {IEEE} Trans. Inf. Theory}, vol. 63, no. 7, pp. 4298--4316, July
  2017.

\bibitem{Sberlo2020}
O.~Sberlo and A.~Shpilka,
\newblock ``On the performance of {Reed{\textendash}Muller} codes with respect
  to random errors and erasures,''
\newblock in {\em Annual ACM-SIAM Symp. on Discrete Algorithms}, 2020, p.
  1357–1376.

\bibitem{Sidel1992}
V.~M. Sidel'nikov and A.~S. Pershakov,
\newblock ``Decoding of {Reed{\textendash}Muller} codes with a large number of
  errors,''
\newblock {\em Problemy Peredachi Informatsii}, vol. 28, pp. 80--94, 1992.

\bibitem{Bertram2013}
J.~Bertram, P.~Hauck, and M.~Huber,
\newblock ``An improved majority-logic decoder offering massively parallel
  decoding for real-time control in embedded systems,''
\newblock {\em {IEEE} Trans. Commun.}, vol. 61, no. 12, pp. 4808--4815, Dec.
  2013.

\bibitem{Hiller2015}
M.~Hiller, L.~Kurzinger, G.~Sigl, S.~Muelich, S.~Puchinger, and M.~Bossert,
\newblock ``Low-area reed decoding in a generalized concatenated code
  construction for {PUFs},''
\newblock in {\em {IEEE} Computer Society Annual Symp. on VLSI}, July 2015.

\bibitem{Dumer2004}
I.~Dumer,
\newblock ``Recursive decoding and its performance for low-rate
  {Reed{\textendash}Muller} codes,''
\newblock {\em {IEEE} Trans. Inf. Theory}, vol. 50, no. 5, pp. 811--823, May
  2004.

\bibitem{Dumer2006}
I.~Dumer and K.~Shabunov,
\newblock ``Soft-decision decoding of {Reed{\textendash}Muller} codes:
  {Recursive lists},''
\newblock {\em {IEEE} Trans. Inf. Theory}, vol. 52, no. 3, pp. 1260--1266, Mar.
  2006.

\bibitem{Hashemi2018}
S.~A. Hashemi, N.~Doan, M.~Mondelli, and W.~J. Gross,
\newblock ``Decoding {Reed{\textendash}Muller} and polar codes by successive
  factor graph permutations,''
\newblock in {\em Int. Symp. on Turbo Codes {\&} Iterative Inf. Proc. (ISTC)},
  Dec. 2018.

\bibitem{Santi2018}
E.~Santi, C.~H\"ager, and H.~D. Pfister,
\newblock ``Decoding {Reed{\textendash}Muller} codes using minimum-weight
  parity checks,''
\newblock in {\em IEEE Int. Symp. Inf. Theory ({ISIT})}. June 2018, {IEEE}.

\bibitem{Sakkour2005}
B.~Sakkour,
\newblock ``Decoding of second order {Reed{\textendash}Muller} codes with a
  large number of errors,''
\newblock in {\em {IEEE} Inf. Theory Workshop}, Oct. 2005.

\bibitem{Thangaraj2020}
A.~Thangaraj and H.~D. Pfister,
\newblock ``Efficient maximum-likelihood decoding of {Reed{\textendash}Muller}
  {RM}(m-3,m) codes,''
\newblock in {\em Int. Symp. Inf. Theory (ISIT)}, June 2020.

\bibitem{Ye2020}
M.~Ye and E.~Abbe,
\newblock ``Recursive projection-aggregation decoding of
  {Reed{\textendash}Muller} codes,''
\newblock {\em {IEEE} Trans. Inf. Theory}, vol. 66, no. 8, pp. 4948--4965, Aug.
  2020.

\bibitem{Beext86}
Y.~Be'ery and J.~Snyders,
\newblock ``Optimal soft decision block decoders based on fast {H}adamard
  transform,''
\newblock {\em {IEEE} Trans. Inf. Theory}, vol. 32, no. 3, pp. 355--364, May
  1986.

\bibitem{Agrawal2015}
A.~Agrawal, R.~Bairathi, and A.~Joshi,
\newblock ``{FPGA} implementation of 4-point and 8-point fast {H}adamard
  transform,''
\newblock {\em International Journal of Computer Applications}, vol. 124, no.
  3, pp. 23--28, Aug. 2015.

\end{thebibliography}
\vfill
\end{document}